\documentclass{PoS}

\usepackage{cite}

\newcommand{\ov}{\overline}
\newcommand{\imag}{{\rm Im}\,}
\newcommand{\real}{{\rm Re}\,}
\newcommand{\eq}[1]{Eq.~(\ref{#1})}
\newcommand{\eqsand}[2]{Eqs.~(\ref{#1}) and (\ref{#2})}
\newcommand{\fig}[1]{Fig.~\ref{#1}}

\newcommand{\tev}{\,\mbox{TeV}}
\newcommand{\gev}{\,\mbox{GeV}}

\newcommand{\CP}{\ensuremath{C\!P}\ }

\newcommand{\Bbar}{\,\overline{\!B}}
\newcommand{\bbq}{$B_q\! - \ov{\!B}{}_q$}
\newcommand{\bbd}{$B_d\! - \ov{\!B}{}_d$}
\newcommand{\bbs}{$B_s\! - \ov{\!B}{}_s$}

\newcommand{\bbms}{\bbs\ mixing}

\newcommand{\bbmq}{\bbq\ mixing}

\newcommand{\dm}{\ensuremath{\Delta m}}
\newcommand{\dg}{\ensuremath{\Delta \Gamma}}
\newcommand{\lt}{\left}
\newcommand{\rt}{\right}

\newcommand{\ket}[1]{\ensuremath{| #1 \rangle }}

\newcommand{\braOket}[3]{\langle#1|#2|#3\rangle}
\newcommand{\epm}[2]{
 \raisebox{-0.5ex}{\shortstack[l]{$\scriptstyle+#1$\\$\scriptstyle-#2$}}}
\newcommand{\ds}{\displaystyle}

\newcommand{\nn}{\nonumber \\}

\title{Penguin pollution in $\beta$ and $\beta_s$}

\ShortTitle{Penguin pollution in $\beta$ and $\beta_s$}

\author{\speaker{Ulrich Nierste}\\
        Institute for Theoretical Particle Physics (TTP)\\
        Karlsruhe Institute of Technology (KIT)         \\
        76131 Karlsruhe, Germany\\ 
        E-mail: \email{ulrich.nierste@kit.edu}}

      \abstract{
        The mixing-induced CP asymmetries in $B_d \to J/\psi K_S$ and
        $B_s \to J/\psi \phi$ are essential to detect or constrain new
        physics in the \bbd\ and \bbms\ amplitudes, respectively. To
        this end one must control the penguin contributions to the decay
        amplitudes, which affect the extraction of fundamental CP phases
        from the measured CP asymmetries. Although the ``penguin
        pollution'' is doubly Cabibbo-suppressed, it could compete in
        size with current experimental errors. In this talk I present 
        a calculation of the penguin contributions treating QCD effects
        with soft-collinear factorisation and compare method and results
        with the alternative approach employing flavour-SU(3) symmetry. 
        As a novel feature, I present results for the 
        penguin pollution in $b\to c\ov c d$ modes.}

\FullConference{9th International Workshop on the CKM Unitarity Triangle\\
                28  November - 3 December 2016\\
                Tata Institute for Fundamental Research (TIFR), Mumbai,
                India}

\begin{document}

\section{Introduction}
\begin{figure}[t]
\centering
\includegraphics[width=0.27\textwidth]{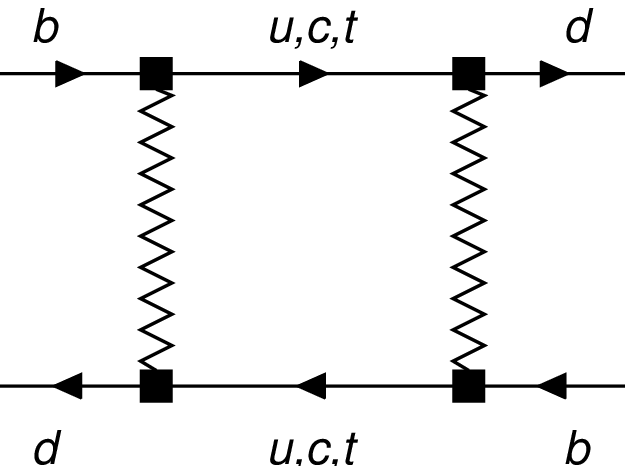}\hspace{3cm}
\includegraphics[width=0.27\textwidth]{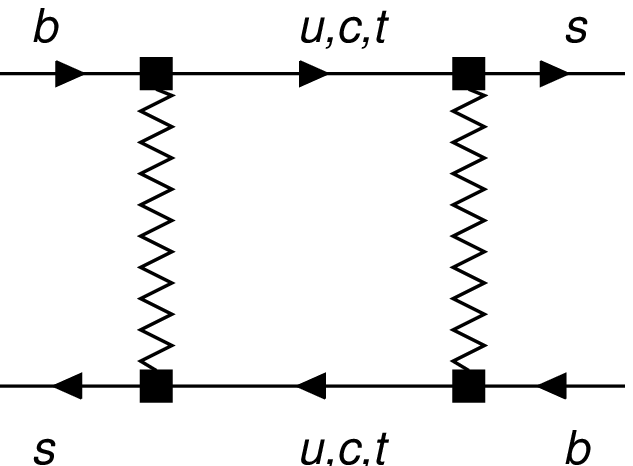}
\caption{Box diagrams describing \bbd\ and \bbms\ in the Standard
  Model.\label{fig:box}}
~\\[-3mm]
\hrule
\end{figure}

In this talk I discuss time-dependent CP asymmetries
\begin{eqnarray}
A_{\rm CP}^{B_q\to f}(t) \equiv
\frac{\Gamma(\Bbar_q(t)\to f)-\Gamma(B_q(t)\to f)}{\Gamma(\Bbar_q(t)\to f)+
\Gamma(B_q(t)\to f)},\qquad\qquad q=d\mbox { or }s, \label{eq:defacp}
\end{eqnarray}
for $B_{d,s}$ decays into final states $f$ consisting of a charmonium
and a light pseudoscalar or vector boson.  Prime examples are the decays
$B_d \to J/\psi K_S$ and $B_s \to J/\psi \phi$, which are both triggered
by the quark decay $\ov b\to \ov cc \ov s$.  I only consider the case
that $f$ is a CP eigenstate; if $f$ comprises two vector mesons (as in
$B_s \to J/\psi \phi$) it is understood that the CP-even and CP-odd
components are properly separated through an angular
analysis. 
Precise
measurements of these mixing-induced CP asymmetries serve to determine
the CP phases related to the \bbd\ and \bbms\ amplitudes. Within the
Standard Model these are
\begin{eqnarray}
  2\beta \equiv \arg\left(
    \frac{V_{tb}V_{td}^*}{V_{cb}V_{cd}^*}\right)^2\qquad  &&\;
  \mbox{and} \qquad
2\beta_s\equiv \arg \left(
    \frac{V_{tb}^*V_{ts}}{V_{cb}^*V_{cs}}\right)^2 . 
\label{eq:cpnum}
\end{eqnarray}
Here $(V_{tb}V_{tq}^*)^2$ stems from the box diagrams shown in
\fig{fig:box}.   The
\bbmq\ amplitudes probe virtual effects of new particles with masses
as high as 100$\tev$, if new physics enters \bbmq\ at tree level. It is
therefore of utmost importance to control the theoretical uncertainties
in the relation between the measured $A_{\rm CP}^{B_q\to f}(t)$ and the
fundamental CP phases in \eq{eq:cpnum} as precisely as possible.

The CP asymmetry in \eq{eq:defacp} reads 
\begin{equation}
A_{\rm CP}^{B_q\to f}(t) =
\frac{S_f \sin(\dm_q t)- C_f \cos(\dm_q t)}{\cosh(\dg_q
  t/2)+A_{f,\dg_q}\sinh(\dg_q t/2)}. \label{eq:acp}
\end{equation}
Here $\dm_q$ and $\dg_q$ are the mass and width difference,
respectively, between the mass eigenstates of the \bbq\ system.  $\dm_q$
and $\dg_q$ are CP-conserving quantities calculated from the box
diagrams in \fig{fig:box}. In $B_d$ decays we can set the denominator in
\eq{eq:acp} to 1, because $\dg_d$ is very small.  The coefficients
$S_f$, $C_f$, and
$A_{f,\dg_q}$ 
depend on the decay amplitude $A(B_q \to f)$. For the $\ov b\to \ov cc
\ov s$ amplitudes of interest one usually writes:
\begin{equation}
A (B_q \to f) = V_{cb}^*V_{cs} T_f + V_{ub}^*V_{us} P_f . \label{eq:atp}
\end{equation}
The ``tree'' and ``penguin'' amplitudes read 
\begin{eqnarray}
T_f &=& \frac{G_F}{\sqrt{2}} \braOket{f}{
     C_1 Q_1^c +C_2 Q_2^c + \sum_{j} C_j Q_j}{B_q},\\
 P_f &=& \frac{G_F}{\sqrt{2}} \braOket{f}{
     C_1 Q_1^u +C_2 Q_2^u + \sum_{j} C_j Q_j}{B_q}. \label{eq:tpc}
\end{eqnarray}
Here $G_F$ is the Fermi constant and $Q_1^q=\ov q{}^\alpha \gamma_\mu
(1-\gamma_5) s^\beta \ov b{}^\beta \gamma^\mu (1-\gamma_5)q^\alpha$ and
$Q_2^q=\ov q{}^\alpha \gamma_\mu (1-\gamma_5) s^\alpha \ov b{}^\beta
\gamma^\mu (1-\gamma_5)q^\beta$ are the current-current operators
generated by $W$-boson exchange. The sum over $j$ comprises the penguin
operators $Q_{3-6}$ and the chromomagnetic operator $Q_{8G}$ (see
Ref.~\cite{bbl} for the definitions). The $C_k$'s are the Wilson
coefficients which encode the short-distance physics; the top-quark
penguin loops (entering $C_{3-6}$ and $C_{8G}$) appear in both $T_f$ and
$P_f$, because the CKM unitarity relation
$V_{tb}^*V_{ts}=-V_{cb}^*V_{cs}-V_{ub}^*V_{us}$ is used to eliminate
$V_{tb}^*V_{ts}$ from \eq{eq:tpc}. Expanding to first order in
$\epsilon=|V_{us}V_{ub}/(V_{cs}V_{cb})|\approx 0.02$ one has
\begin{eqnarray}
S_f \simeq -\eta_f \sin(\phi_q+ \Delta \phi_q) &&\qquad\quad
\mbox{with~~~~~~~}    
 \tan(\Delta \phi_q) \simeq  \; 2
    \epsilon\sin \gamma \; \real \frac{P_f}{T_f} ,
        \label{eq:Dphi}
\end{eqnarray}
where $\CP\!\ket f=\eta_f\ket f$ with $\eta_f=\pm 1$, $\phi_d=2\beta$,
and $\phi_s=-2\beta_s$. Furthermore,  $C_f\simeq 2 \epsilon\sin\gamma\, \imag (P_f/T_f)$
quantifies direct CP violation.

$\Delta \phi_q$ in \eq{eq:Dphi} is the \emph{penguin pollution} which
obscures a clean extraction of $\phi_q$ from the measured $S_f$.  The
size of the penguin pollution depends on the considered decay mode
through $\real (P_f/T_f)$ in \eq{eq:Dphi}. A standard way to estimate
$\Delta \phi_q$ employs the flavour-SU(3) symmetry of QCD or its SU(2)
subgroup U-spin. The latter connects pairs of hadronic matrix elements
related by the interchange of down and strange quarks. In the case of
$B_d\to J/\psi K_S$ one can extract the desired $P_f/T_f$ from control
channels such as $B_s\to J/\psi K_S$ or $B_d\to J/\psi \pi^0$, which are
induced by the quark decay $\ov b\to \ov cc \ov d$. In these control
channels the CKM factor $\epsilon$ is replaced by
$|V_{ud}V_{ub}/(V_{cd}V_{cb})|\approx 0.38$ which permits to determine
$P_f/T_f$ from the coefficients $C_f$ and $S_f$ measured in these
modes. In this way one finds the values $-3.9^\circ \leq \Delta\phi_d
\leq -0.8^\circ $ \cite{Faller:2008zc}, $|\Delta\phi_d| \leq 1.6^\circ$
\cite{Ciuchini:2011kd}, $|\Delta\phi_d| \leq 0.8^\circ $
\cite{Jung:2012mp}, and $\Delta\phi_d= -1.1^\circ
\epm{0.85^\circ}{0.7^\circ}$ \cite{DeBruyn:2014oga} for $f=J/\psi K_S$.
The values (listed in chronological order) become more accurate with
more precise data the on the control channels. A general drawback of the
method is the unknown size of SU(3)$_f$ breaking caused by unequal
strange and down quark masses.  SU(3)$_f$ symmetry can be very accurate,
as e.g. in semileptonic $B_{d,s}$ decays, but may also fail completely:
for example, a $b$ quark fragments into a $B_d$ meson almost four times
more often than into a $B_s$. In the case of $B_s \to J/\psi \phi$ one
faces the problem that the $\phi$ meson is an equal mixture of an octet
and a singlet of SU(3)$_f$ symmetry. It is not clear how to treat
SU(3)$_f$ breaking in such a case of maximal symmetry violation and the
method may fail in this case.

The experimental world average $ 2\beta + \Delta \phi_d = 43.8^\circ\pm
1.4^\circ $ \cite{hfag} is 
dominated by $B_d\to J/\psi K_S$, so that $\Delta \phi_d$ here can be
identified with the penguin pollution in this mode. The experimental
error is comparable in size with the expected penguin pollution. The
situation is similar with the experimental value $2 \beta_s+\Delta
\phi_s=1.7^\circ\pm 1.9^\circ$ \cite{hfag} which dominantly stems from
LHCb data on $B_s\to J/\psi K^+ K^-$ and $B_s\to J/\psi f_0[\to
\pi^+\pi^-]$, with an experimental error of $2.2^\circ$ on $2
\beta_s+\Delta \phi_s$ \cite{lhcb}. The statistical powers of $B_s\to
J/\psi \phi [\to K^+ K^-]$, non-resonant $B_s\to J/\psi K^+ K^-$, and
$B_s\to J/\psi f_0[\to \pi^+\pi^-]$ on the determination of $2
\beta_s+\Delta \phi_s$ are 52\%, 8\%, and 42\%, respectively
\cite{stephie}.  The value of $2 \beta_s$ inferred from a global fit to
the CKM unitarity triangle is $2 \beta_s= 2.12^\circ \pm 0.04^\circ$
\cite{ckmf}.

In this talk I present calculations of the penguin contributions to CP
asymmetries which do not use SU(3)$_F$ symmetry, but instead employ 
soft-collinear factorisation in QCD
\cite{Frings:2015eva}.

\section{Operator Product Expansion}
\begin{figure}[t]
\centering \vspace{-3em}
\includegraphics[width=0.35\textwidth]{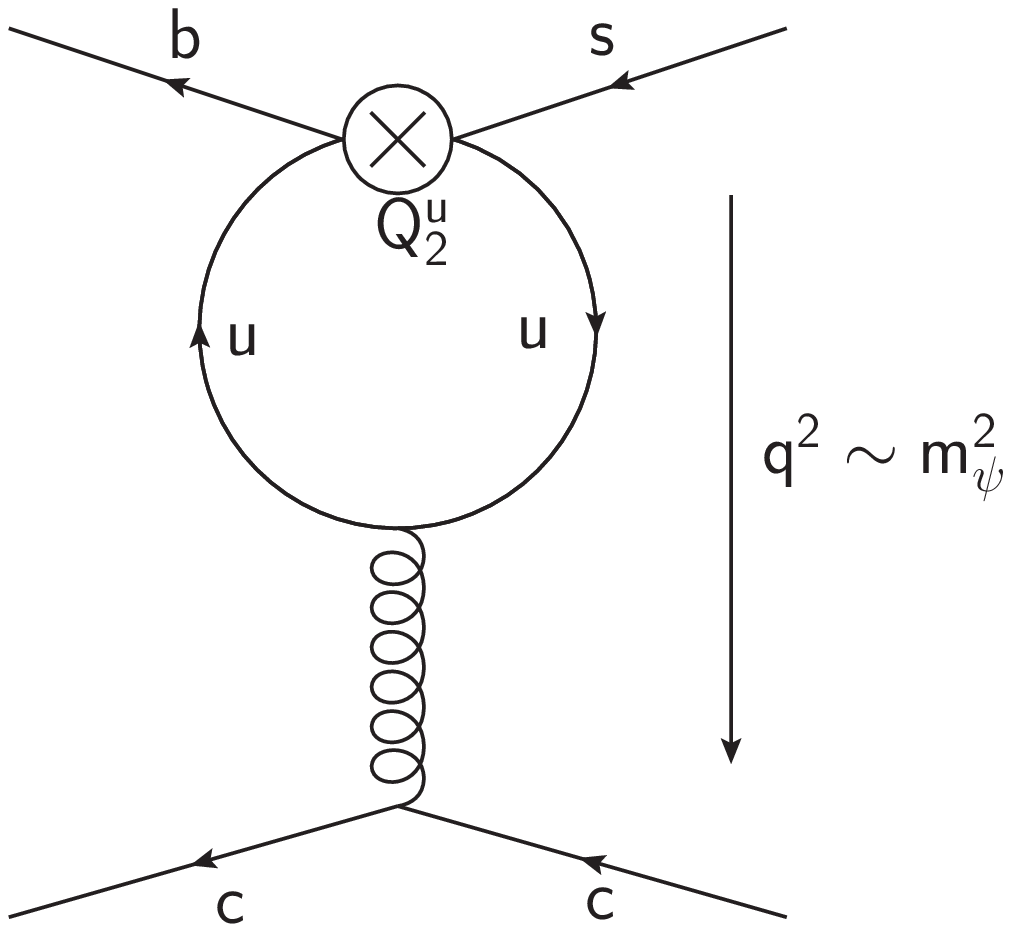}
        ~~~~
\parbox[b]{0.16\textwidth}{\mbox~\\[3cm]
                  $$q^2 \gg \Lambda_{QCD}^2 $$ \mbox~\\[-8mm]
                  \mbox~~~~~~~\scalebox{1.3}{$\ds \longrightarrow $}\\[13mm]}
        ~~~~
\parbox[b]{0.35\textwidth}{\includegraphics[
  width=0.35\textwidth]{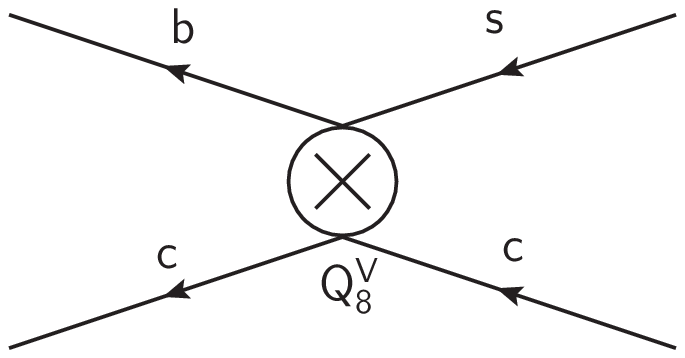}\\[2mm]}
\caption{Pictorial representation of the OPE for the up-quark loop:
Since the momentum transfer $q$ to the charmonium is large, we can 
express the left diagram as the product of a perturbative Wilson
coefficient and the effective four-quark operator on the right. 
\label{fig:ope}}
~\\[-3mm]
\hrule
\end{figure}

\begin{figure*}[t]
                
\centering
                \includegraphics[height=0.2\textwidth,clip=true]%
                  {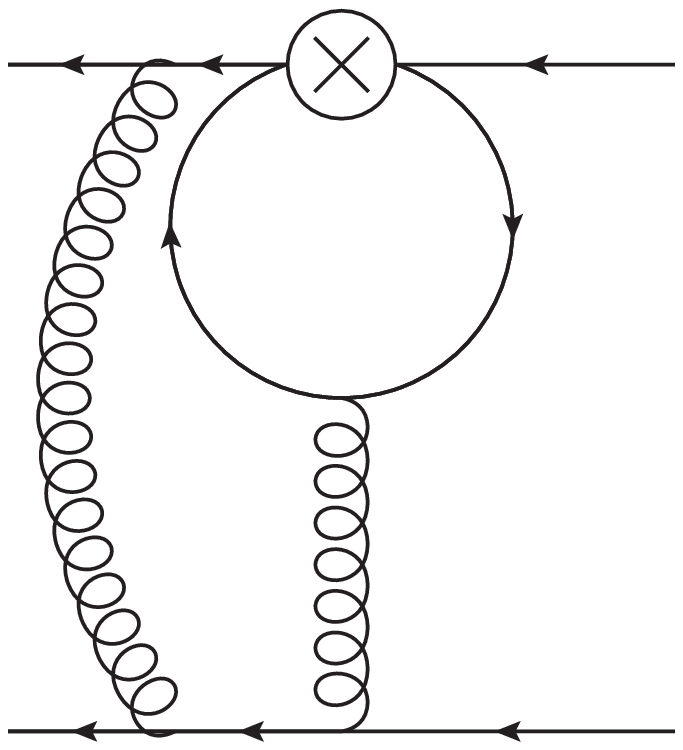} \hspace{10mm}
                \includegraphics[height=0.18\textwidth,clip=true]%
                  {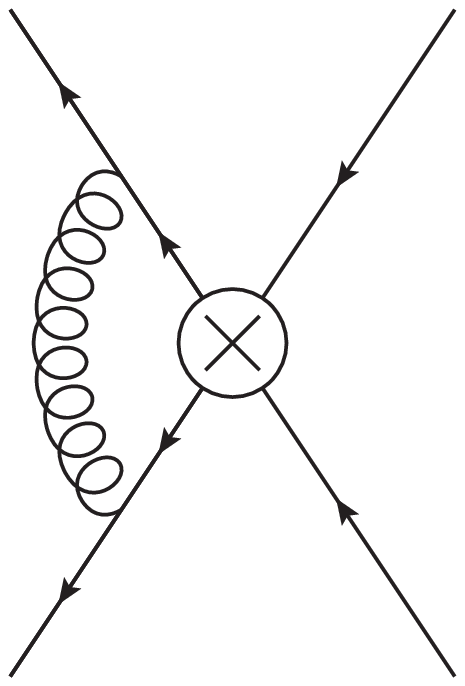} \hspace{10mm}
                \includegraphics[height=0.2\textwidth,clip=true]%
                   {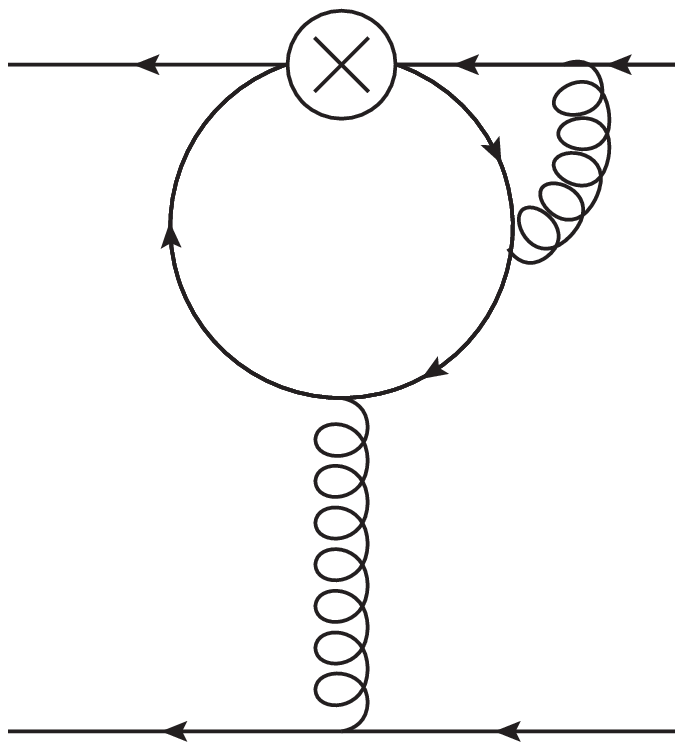} \hspace{10mm}
                \includegraphics[height=0.2\textwidth,clip=true]%
                      {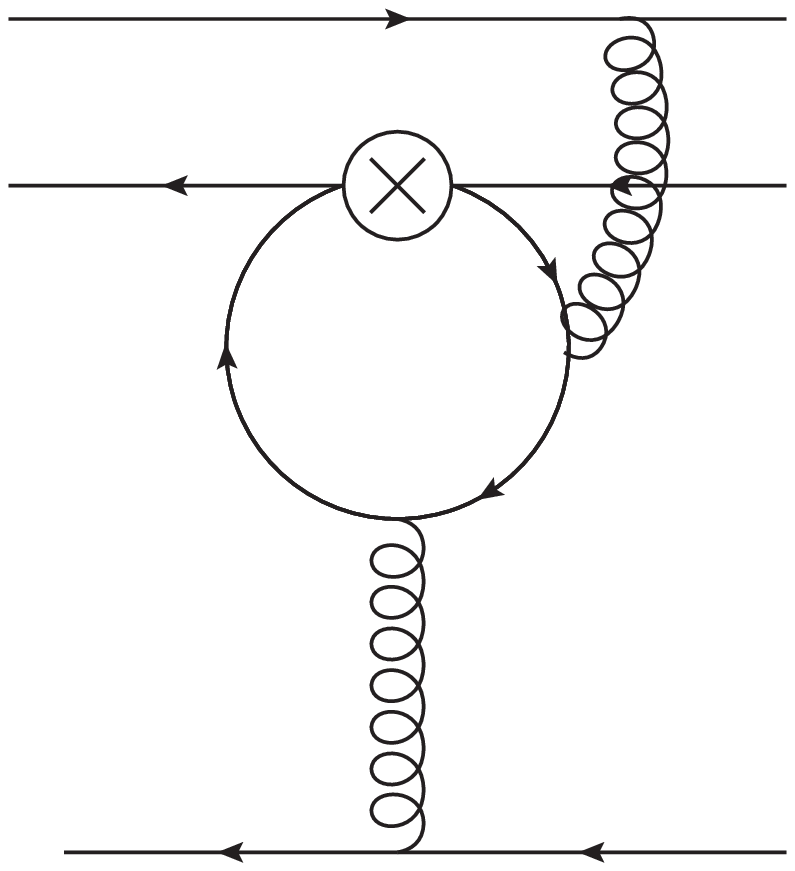}

  \caption{The soft IR divergence of the
    first diagram (contributing to the LHS of \eq{eq:ope}) 
    factorises with the corresponding diagram 
    of the local operator (RHS of \eq{eq:ope}) shown next.  
    The third diagram has a collinear IR 
    divergence and finally a spectator-scattering diagram is 
    shown.\label{fig:pengs}}
  ~\\[-3mm]
  \hrule
\end{figure*}

Many physical problems involve a hard scale $\sqrt{q^2}$ which is much
larger than the fundamental scale $\Lambda_{\rm QCD}\sim 0.4\gev $ of
QCD.  The operator product expansion (OPE) is a calculational tool to
express the quantity of interest in terms of a series in $\Lambda_{\rm
  QCD}/\sqrt{q}$. In our case we a apply the OPE to $P_f$ in \eq{eq:tpc}
and $\sqrt{q^2}\sim m_{\psi}\sim 3\gev$ is the hard scale. The
troublesome contribution to $P_f$ stems from $Q_{1,2}^u$ in \eq{eq:tpc};
the corresponding one-loop contribution is shown in \fig{fig:ope}.
The OPE for the contribution of $Q_j^u$, $j=1,2$, to $P_f$ for
$B_d \to J/\psi K_S$ reads
\begin{eqnarray}
  \!\!  \braOket{J/\psi K_S}{Q_j^u}{B_d} &=&  \sum_k
  \widetilde{C}_{j,k}
  \braOket{J/\psi K_S}{Q_{k}}{B_{d}}
    + \ldots 
\label{eq:ope}
\end{eqnarray}
Here $k=0V,0A,8V,8A$ labels different local four-quark operators
with flavour structure $\ov b s\, \ov c c$:
 \begin{eqnarray}
Q_{0V} & \equiv & \ov b{} \gamma_\mu (1-\gamma_5) s \,
                  \ov c{}  \gamma^\mu c ,
                \nn
Q_{0A} & \equiv & \ov b{} \gamma_\mu (1-\gamma_5) s \,
                  \ov c{}  \gamma^\mu \gamma_5 c,
\nn
Q_{8V} & \equiv & \ov b{} \gamma_\mu (1-\gamma_5) T^a s\,
                  \ov c{}  \gamma^\mu T^a c,
                \nn
Q_{8A} & \equiv & \ov b{} \gamma_\mu (1-\gamma_5) T^a s\,
                  \ov c{}  \gamma^\mu \gamma_5 T^a c.
\label{eq:ops}
\end{eqnarray}
These operators suffice to reproduce $P_f$ at the leading power of
$\Lambda_{\rm QCD}/\sqrt{q}$. Sub-leading powers involve additional
operators , which are indicated by the dots in \eq{eq:ope}.  The Wilson
coefficients $\widetilde{C}_{j,k}$ in \eq{eq:ope} are found by
calculating $\ov b\to \ov c c \ov s $ Feynman diagrams with $Q_j^u$ in
the desired order of $\alpha_s$ and comparing the result with Feynman
diagrams involving the operators $Q_{k}$ in \eq{eq:ops} in the
corresponding order of QCD. The leading non-vanishing order, shown in
\fig{fig:ope}, only involves the operator $Q_{8V}$ on the right-hand
side (RHS) of the OPE in \eq{eq:ope}. 
The coefficient is
\begin{eqnarray}
\widetilde{C}_{2,8V}^{(0)}
& = &\frac23 \frac{\alpha_s}{4 \pi}\lt[ \ln\lt(\frac{q^2}{\mu^2}\rt) -i 
\pi -\frac23 \rt], \label{eq:pf}
\end{eqnarray}
where $\mu$ is the renormalisation scale.  The idea to factorise the
one-loop diagram in this way was proposed by Bander, Silverman, and Soni
(BSS) in Ref.~\cite{Bander:1979px} and applied to $B_d\to J/\psi K_S$ in
Ref.~\cite{Boos:2004xp}. In order to establish the OPE in \eq{eq:ope}
one must prove that the coefficients $\widetilde{C}_{j,k}$ are free from
infrared singularities, which involves the study of higher orders in
$\alpha_s$. This proof has been carried out in
Ref.~\cite{Frings:2015eva} and involves the analysis of (i) soft IR
divergences of the two-loop diagrams contributing to $\langle
Q_j^u\rangle$, (ii) collinear IR divergences of these diagrams, (iii)
spectator scattering diagrams, and (iv) higher-order diagrams in which
the large momentum bypasses the penguin loop (``long distance
penguins''). Sample diagrams are shown in \fig{fig:pengs}.
In Ref.~\cite{Frings:2015eva} it has been shown that indeed all infrared
singularities properly factorise and cancel from the coefficients
$\widetilde{C}_{j,k}$, which therefore can be calculated perturbatively
order-by-order in $\alpha_s$. The leading order (LO) contribution to the
$\widetilde{C}_{j,k}$ stems from the penguin operators $Q_{3-6}$ in
\eq{eq:tpc}, which contribute trivially to \eq{eq:ope} as local $\ov b
s\, \ov c c$ operators. The dependence of $C_{3-6}$ on the unphysical
renormalisation scale $\mu$ cancels (to order $\alpha_s$) with the
$\mu$-dependent terms of the next-to-leading order (NLO)
corrections. The result in \eq{eq:pf} belongs to the NLO and depends on
the renormalisation scale and scheme. It is only meaningful in
combination with the LO contributions involving $C_{3-6}$, so that these
scale and scheme dependences cancel. In Ref.~\cite{Boos:2004xp} the LO
contribution has been omitted and the inferred penguin pollution is
substantially smaller than the one found by us.

The standard application of soft-collinear factorisation in flavour
physics addresses $B$ decays into two light mesons (QCD factorisation)
\cite{qcdf}. In our case instead one of the final-state mesons is heavy
and the $J/\psi$ mass is the relevant heavy scale in the problem. As a
consequence, we cannot factorise the matrix element of colour-octet
four-quark operators into a form factor and a decay constant \cite{nofac}.

\section{Matrix elements and numerical results}
In order to predict the size of the penguin pollution $\Delta \phi_d$
for $B_d\to J/\psi K_S$ in \eq{eq:Dphi} from the calculated $\real
(P_f/T_f)$ we need the (ratios of the) hadronic matrix elements 
\begin{eqnarray}
v_0 V_0 \equiv \braOket{J/\psi K^0}{Q_{0V}}{B_{d}}, &&\qquad  
a_0 V_0 \equiv \braOket{J/\psi K^0}{Q_{0A}}{B_{d}}, \nn 
v_8 V_0 \equiv \braOket{J/\psi K^0}{Q_{8V}}{B_{d}}, &&\qquad  
a_8 V_0 \equiv \braOket{J/\psi K^0}{Q_{8A}}{B_{d}}. \label{eq:rme} 
\end{eqnarray}
\eq{eq:rme} defines the complex parameters $v_{0,8}$ and $a_{0,8}$ with
the common normalisation factor
\begin{eqnarray}
  V_0 \equiv\; \braOket{J/\psi K^0}{Q_{0V}}{B_{d}}_{\rm fact}
  =\; 2 f_{J/\psi} m_{B_d} p_{cm} F_1^{B\to K}(m_{\psi}^2)=\; (4.26 \pm 0.16)
  \gev{}^{3}. \label{eq:fac}
\end{eqnarray}
$V_0$ is the factorised matrix element of the colour-singlet operator
$Q_{0V}$ involving the $J/\psi$ decay constant $ f_{J/\psi}$, the $B_d$
mass $m_{B_d}$, the magnitude of the $K_S$ center-of mass three-momentum
$p_{cm}$, and the form factor $F_1^{B\to K}$. Next $v_{0,8}$ and
$a_{0,8}$ are categorised in terms of $1/N_c$ counting, where $N_c=3$ is
the number of colours. One has $v_0=1+{\cal O} (1/N_c^2) $, $v_8, a_8 =
{\cal O} (1/N_c) $, and $a_0={\cal O} (1/N_c^2) $. It is well-known that
the coefficient of $v_0$ in $T_f$ is small, so that the branching ratio
$B(B_d\to J/\psi K_S)$ is dominated by $v_8$ and $a_8$. Therefore we can
use the measured $B(B_d\to J/\psi K_S)_{\rm exp}$ as a cross-check of
our colour counting for the peculiar colour-octet matrix elements. With
the numerical values of the Wilson coefficients and \eq{eq:fac} one
finds \cite{Frings:2015eva}:
\begin{eqnarray}
  \frac{B(B_d\to J/\psi K_S)}{B(B_d\to J/\psi K_S)_{\rm exp}}
&=& 
  \lt[1\pm 0.08 \rt] \lt| 0.47v_0 +7.8 (v_8-a_8) \rt|^2.
\label{eq:bf}
\end{eqnarray}
This implies $0.07\leq |v_8-a_8|\leq 0.19$ if $v_0$ is set to 1,
illustrating that the colour counting works for the branching
ratio. $a_0$ comes with small coefficients in both $T_f$ and $P_f$ and
is negligible. For the prediction of $P_f/T_f$ at NLO we need $v_8$
and impose $|v_8|\leq 1/3$, complying with colour counting, and vary 
the phases of the matrix elements between $-\pi$ and $\pi$. The result
is \cite{Frings:2015eva}
\begin{equation}
 |\Delta_{d}| \leq 0.68^\circ,\qquad \quad
 |C_{J/\psi K_S}| \leq 1.33 \cdot 10^{-2}.   
\end{equation}
The bound on $\Delta_{d}$ is comparable to the one derived from SU(3)$_F$
symmetry (quoted in the introduction), but sharper. 

In the case of $B_s\to J/\psi \phi$ one finds
\begin{center}
\begin{tabular}{c@{~~~~~}c@{~~~~~}c}
  $\ds (J/\psi\phi)^0$&  $\ds (J/\psi\phi)^\parallel$& $\ds
  (J/\psi\phi )^\perp$
  \\\hline &&\\[-3mm] 
  $\ds |\Delta \phi_s|\leq 0.97^\circ$ & $\ds |\Delta
  \phi_s|\leq1.22^\circ$ & 
  $\ds |\Delta \phi_s|\leq0.99^\circ $ \\[1mm]
  $\ds |C_f| \leq 1.89\cdot 10^{-2}$ & $\ds |C_f| \leq 2.35\cdot
  10^{-2}$ & 
  $\ds |C_f| \leq 1.92\cdot 10^{-2}$ 
\end{tabular}
\end{center}
for the scalar, parallel, and perpendicular polarisation states, respectively.
 
As a novel feature, the method of Ref.~\cite{Frings:2015eva} permits the
prediction of the penguin contributions to $\ov b \to \ov c c \ov d$
decays, for example:
\begin{eqnarray}
  B_d\to J/\psi \pi^0: \qquad && \quad|S_{J/\psi \pi^0} + \sin(2\beta)| \leq 0.18, \qquad 
  |C_{J/\psi \pi^0}| \leq 0.29. \qquad \label{eq:pi0} \\
  B_s \to J/\psi K_S: \qquad &&  |S_{J/\psi K_S} - \sin(-2\beta_s)| \leq 0.26,
  \qquad                     |C_{J/\psi K_S}| \leq 0.27. \qquad \label{eq:bsc}
\end{eqnarray}
The first result means $-0.86\leq S_{J/\psi \pi^0} \leq
-0.50$. \eq{eq:pi0} favours the Belle result \cite{Lee:2007wd}
$S_{J/\psi \pi^0}=-0.67\pm 0.22$, $C_{J/\psi \pi^0}=-0.08\pm 0.17$ over
the BaBar result \cite{Aubert:2008bs} $S_{J/\psi \pi^0}=-1.23\pm 0.21$,
$C_{J/\psi \pi^0}=-0.20\pm 0.19$. Predictions for more $\ov b \to \ov c
c\ov s$ and $\ov b \to \ov c c\ov d$ modes can be found in Tab.~1 of
Ref.~\cite{Frings:2015eva}. 

It is worthwhile to compare the methods and results presented in this
talk with those of the alternative approach based on SU(3)$_F$ symmetry:
It is gratifying to see that the two completely different methods give
compatible results for $\Delta \phi_d$ in the case of $B_d \to J/\psi
K_S$. However, the SU(3)$_F$ estimate of $\Delta \phi_d$ depends on the
choice for the size of SU(3)$_F$ breaking added to the value of
$P_f/T_f$ extracted from the $\ov b \to \ov c c \ov d$ control
channels. In analyses of branching fractions (which probe $T_f$ with
little sensitivity to $P_f$) it is possible to include linear SU(3)$_F$
breaking in the $T_f$ amplitudes and thereby test the quality of the
method from the data (see Ref.~\cite{Jung:2012mp} for $B\to J/\psi X$
decays and Refs.~\cite{Gronau:1995hm} and \cite{Muller:2015lua} for
$B,D$ decays to two light mesons, respectively). However, in the case of
the up-quark loop in $P_f$ there is not enough information to
disentangle the penguin pollution from the matrix elements parametrising
SU(3)$_F$ breaking, no matter how many control channels are included:
The SU(3)$_F$ breaking stemming from the $d\to s$ replacement when
linking the $\ov b \to \ov c c \ov d$ control channel to the $\ov b \to
\ov c c \ov s$ signal process is never constrained by any of these
control channel processes.  On the contrary, the OPE-based approach of
Ref.~\cite{Frings:2015eva} makes enough redundant predictions to
simultaneously test the method and to constrain the penguin pollution
in the $\ov b \to \ov c c \ov s$ decays: Here the litmus test are the
predictions for the various $\ov b \to \ov c c \ov d$ channels (such as
those in \eqsand{eq:pi0}{eq:bsc}), which make the method falsifiable.

The SU(3)$_F$ method utilises the feature that the SU(3)$_F$ symmetry is
approximately exact, with corrections treatable as small (i.e.\ ${\cal
  O}(30\% )$) perturbations. The quality of the symmetry allows us to
assign exact or approximate SU(3)$_F$ quantum numbers to the particle
states, as we routinely do for the light pseudoscalar mesons. In the
case of $B_s\to J/\psi \phi$ one faces the fact that the $\phi$ meson is
an equal mixture of octet and singlet, so that it does not correspond to
an approximate SU(3)$_F$ eigenstate.  There are two possible
explanations of this observations: (i) SU(3)$_F$ is not a good symmetry
for decays into final states with vector mesons. (ii) SU(3)$_F$ breaking
is small, but the spectrum of the ``unperturbed'' strong hamiltonian
(corresponding to the limit $m_s=m_d=m_u$) is almost degenerate, so
that even a small perturbation can lead to maximal mixing. If case (i)
is realised in nature, SU(3)$_F$ cannot be applied to constrain the
penguin pollution in $B_s\to J/\psi \phi$. If (ii) is the correct
explanation, a necessary ingredient of an SU(3)$_F$-based assessment of
the penguin pollution is the determination of both the octet and singlet
matrix elements from the control channels. In addition, one must develop
a formalism which permits the treatment of SU(3)$_F$ breaking for the
case that the final states of the considered decays cannot be
approximated by SU(3)$_F$ eigenstates. In view of this situation it is
safe to say that SU(3)$_F$-based estimates of the penguin pollution in
$B_s\to J/\psi \phi$ rest on shaky ground.

\section{Conclusions and outlook}
In this talk I have presented results of Ref.~\cite{Frings:2015eva} for
the penguin pollution affecting the extractions of the CP phases
$2\beta$ and $2\beta_s$ from the decays $B_d\to J/\psi K_S$ and $B_s\to
J/\psi \phi$, respectively. The predictions are based on a new
calculational approach which utilises an operator product expansion
(OPE) for the penguin amplitude. To establish the OPE the infrared
safety of the Wilson coefficients calculated from the up-quark loop
contribution to the penguin amplitude had to be proven, which elevates
the BSS approach of Ref.~\cite{Bander:1979px} to a field-theoretic
concept applicable at any order of $\alpha_s$. (However, we found no
justification to apply the OPE to the charm-quark loop, which in our
framework resides in the hadronic matrix elements.)  Our method can also
be applied to CP asymmetries in $\ov b \to \ov c c \ov d$ decays, in
which the penguin-to-tree ratio is much larger. As examples I have
quoted bounds on the penguin contributions for the CP asymmetries in the
decays $B_d \to J/\psi \pi^0$ and $B_s \to J/\psi K_S$. The
confrontation of our predictions for $\ov b \to \ov c c \ov d$ decays
with more precise data will be a stringent test of the OPE-based
approach. In the future the errors of the predictions may shrink, if
effort is put into the calculation of the hadronic parameter $v_8$,
possibly with the help of QCD sum rules.  In my talk I have further
expressed a critical view of the application of SU(3)$_F$ symmetry to
the penguin pollution in $B_s\to J/\psi \phi$.

\section*{Acknowledgements}
I thank the organisers for inviting me to this talk.  I am grateful to
Philipp Frings and Martin Wiebusch for a very enjoyable collaboration.
The presented work is supported by BMBF under contract 05H15VKKB1.

\end{document}